# IoT System Case Study: Personal Office Energy Monitor (POEM)

Milan Milenkovic

IoTsense LLC, Dublin, CA, USA

milan@iotsense.com

**Abstract**

This paper describes the design, implementation, and user evaluation of an IoT project focused on monitoring and management of user comfort and energy usage in office buildings. The objective is to depict an instructive use case and to illustrate experiences with all major phases of designing and running a fairly complex IoT system. The design part includes motivation and outline of the problem statement, the resulting definition of data to be collected, system implementation, and subsequent changes resulting from the additional insights that it provided. The user experience part describes quantitative findings as well as key results of the extensive human factors study with over 70 office users participating in two major pilots in France and Japan.

The original idea for this project came out from a diverse group of companies exploring challenges of designing and operating smart buildings with net-positive energy balance. Members included companies involved in the design and construction of smart buildings, building-management and automation systems, computer design, energy systems, and office furniture and space design. One of the early insights was that maximum energy efficiency in office buildings cannot be achieved and sustained without the awareness and active participation of building occupants. The resulting project explored and evaluated several ways to engage and empower users in ways that benefit them and makes them the willing and active participants.

## 1 MOTIVATION AND OBJECTIVES

The project goals were defined by the Net-Positive Energy Consortium formed to explore the impact and synergies of technologies that could be deployed in constructing and operating buildings that are energy net positive, i.e. produce more energy than they consume. This is a challenging and important task since buildings consume approximately 40% of the total energy used in the USA [1]. The international consortium, now dissolved after achieving its stated objectives, included eight major global corporations engaged in all phases of the design, construction and operation of smart buildings. Member companies had product focus and expertise in building construction, building-management systems (two), office furniture and space design, cafeteria management, solar division of an energy production company, computer equipment and printer manufacturing.

One of the key conclusions of the member companies in the problem analysis phase was that the operation of energy-positive buildings will require awareness and active and willing participation of its occupants. Some internal studies projected that informed modifications of user behavior could save 12% of the total energy consumption of the building. It was also observed that in the current practice there is no direct connection between users and the systems that manage the building. As one BMS vendor participant observed "BMS is supposed to manage the comfort and safety of building's occupants, but it has no idea where they are, what they are doing, and what their needs and preferences are".

This is a major missed opportunity, since maintaining the comfort of building occupants is the primary target of a building management system. Moreover, the users - by virtue of their actions and behaviors - are the primary driver of building's energy consumption.

The primary design focus of the project was chosen to be providing of the personalized information to users on their ambient conditions, and creation of a mechanism for communicating their comfort preferences to the management system. A secondary objective was to measure and assess the aggregate and personal energy consumed by the IT equipment since no reliable data on that were available to the consortium to guide the smart building design and consumption projections. The specific initial focus was on the user's personal computer and printer usage. The project was named POEM for Personal Energy Office Monitor. Its focus was on monitoring and managing the demand side, with informed user involvement to minimize consumption, while maintaining user satisfaction, comfort and safety. The supply side, i.e. adding energy production capabilities to achieve the net-positive energy goal, was part of another project not described here.

## 2 POEM SYSTEM DESIGN

Based on the stated objectives and some supporting research [2, 3, 4, 5, 6, 7], the design approach was to implement the pervasive personalized sensing of the user ambient and IT energy usage. An early original slide of the conceptual design of the system to address these objectives is depicted in Figure 1.

The plan was to collect electricity usage data by installing smart (internet enabled) plug load meters in the user's workplace. A somewhat uncommon choice was to attach the ambient sensors for the light temperature and humidity to user's PCs. This is convenient because the PC provides the power, connectivity, and added computational capability to the attached sensor. The resulting combination circumvents the power and connectivity constraints of using wireless sensors. Moreover, users spend a lot of time near their PCs, thus making them a good place to measure personal ambient conditions. On the pragmatic side, one of the companies involved in the project was a computer company with interest in testing the use of sensors on PCs in an office environment.

The plan was to collect sensor data in a sensor database that would interface to the BMS system to report energy consumption and user preferences for ambient settings. The sensor database was also intended to provide data for the personalized visualizations of measurements of interest to individual users and the aggregate reports to system managers.

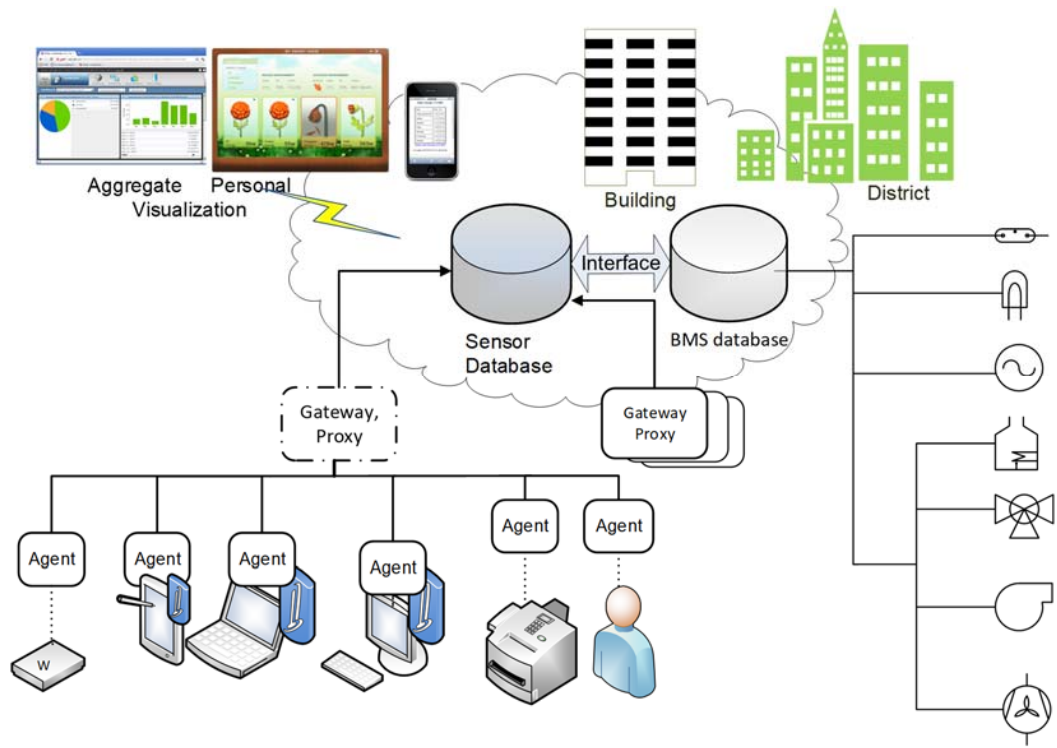

Fig. 1 POEM System Conceptual Design

## 2.1 POEM User Interface Design

In order to define the system functional requirements, project designers started with outlining functions that need to be provided to users. Much of the attention was placed on the user interface design of POEM, both in terms of functionality and appearance. This took several months and intentionally preceded completion of the system architecture in order to make sure that it could provide all the desired functions. The user interface design went through many iterations, starting with whiteboard functional sketches, until it finally settled on the representation depicted in Figure 2.

The POEM UI screen contains fields for user energy consumption feedback, ambient conditions dashboard, personal comfort feedback and notifications. Energy data are represented both in quantitative terms and as intuitive visual indication.

Research [8] and our preliminary studies [9] indicated that users prefer to have an intuitive visual indication of how they are doing, as opposed to a numeric-only indication. The latter imposes the cognitive load of knowing what the good values are supposed to be, which is not always intuitive. For example, for the PC power consumption it was not clear what is the magnitude of the impact of user behavior and usage modality – such as proactively initiating a sleep state when not in use – versus the baseline consumption dictated by the specific configuration and Energy Star rating of their unit [10]. Thus, a user dutifully following energy-saving recommendations could be consuming more energy in comparison with another one who is not if their respective PCs differ significantly in terms of their inherent energy efficiency. In order to encourage good behaviors, the POEM personalized indicators and feedback had to be adjusted accordingly, without burdening the users with the raw data that they would have to scale as appropriate for their particular PC.

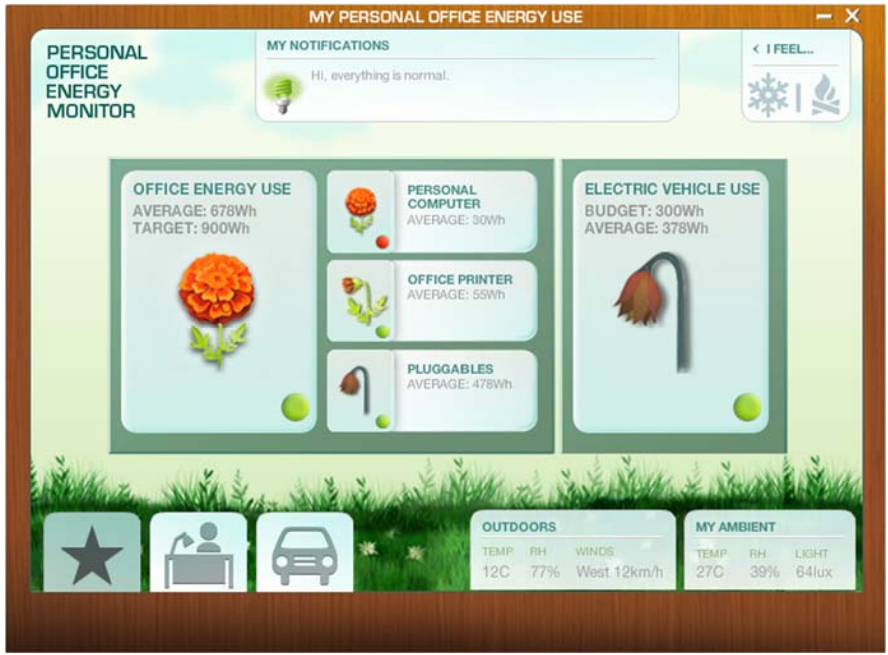

Fig. 2 POEM User Interface, Main (Home) Screen

To deal with these issues, we opted for using a visual metaphor to depict the basic "bad, neutral, good" ratings. In our design we opted to use the shape a flower with a flourishing image depicting the good and a withering flower corresponding to what was in effect the bad state. We renamed that state to something meaning "could use improvement" in order to strive for the positive reinforcement and not to discourage users. We found that the use of a visual metaphor works well and resonates with users. The specific choice of a visual representation, such as a flower, is less important and it is a matter of preference.

Other key information presented on the POEM's home UI screen included the user's personal ambient measurements of temperature, humidity, and light intensity "where I am right now", and outdoors weather for the exact location of their building obtained form an Internet-based service. A separate field labeled "I Feel" allowed users to input their personal subjective feeling of comfort using the ASHRAE seven-point thermal scale (-3 very cold to +3 very hot) [11]. This information was aggregated, annotated with the identifier of the heating and cooling zone from which the user is reporting, and sent to the building manager as a summary report and optionally as a notification, depending on severity. The plan was to later interface this input with the BMS for automated response to mediated user preferences in zones that can be individually controlled.

The field labeled "My Notifications" provides a mechanism for individual and group notifications of users from system operators and building managers, such as ambient adjustments in progress or estimated times for service restoration in case of failures. It can also be used for bolstering energy-saving efforts and campaigns to achieve specific individual and group goals, as well as for implementing challenges and competitions.

Figure 2 depicts the POEM's initial or home UI screen. It allows for user selection of additional screens, such as the electric vehicle charging and management. Individual energy fields are also interactive (clickable) to show details of various aspects of energy consumption. One such screen for office energy totals and comparisons is depicted in Figure 3.

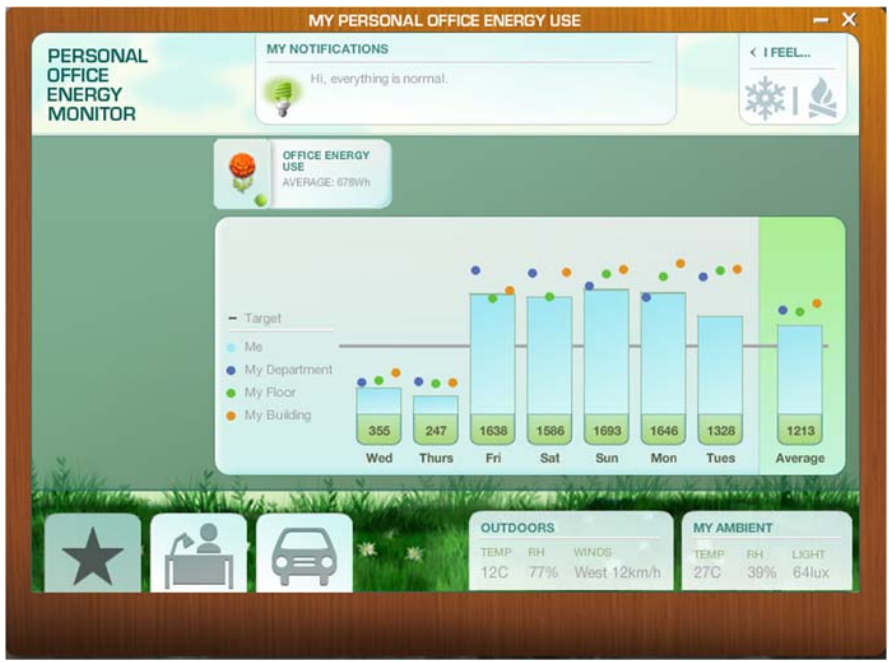

Fig. 3 POEM Energy Usage and Comparison Screen

It shows a user's daily energy consumption for the past week and the computed average that is shown on the main screen. Various colored dots show energy consumption and comparable averages of interest to the user – user's department, floor and building. Our preliminary studies indicated that users can relate to data much better when provided with a point of reference, indicating how their consumption compares to others that may be relevant. As an intuitive example, knowing that you printed 320 pages last month is of some value, but knowing that the average for your peers in the department was 130 gives you a much more useful and perhaps actionable frame of reference. User anonymity and privacy in such reports can be preserved by showing the individual data only to the originating user, and only group ranges or averages for everything else. The horizontal grey line indicates the personalized target objective that the user is aiming for.

The visual indication of user's progress, depicted by the shape of the flower, was computed based on the difference between the target and the actual consumption. Target values could be assigned by users or by the system, depending on the policies in effect. Our strong preference was for the user-settable targets with the management policy settings as a fallback for users who were confused by options or did not want to make a choice themselves.

## 2.2 POEM System Architecture and Implementation

The focus on personalized user feedback and functionality drove many of the system design requirements. In addition to the typical IoT sensor data acquisition and processing, it was necessary to support attribution of measurements to individual points of origin, users, and groups of users determined by POEM categories. This was accomplished by the extensive use of metadata to annotate inputs and coordinated handling of data and metadata at the backend to maintain the required associations.

Other major aspects of the conceptual design included placement of sensors near users, on their PCs for ambient sensing and instrumented power plugs for tracking their energy consumption. Edge nodes were collecting sensor data and providing the user interface, with most of the data processing and storage delegated to the back-end servers. Connectivity and security were basically provided by the network connections and protocol stacks of the host PCs, and the corporate IT mechanisms and policies provided by the enterprise networks on which they were installed. Data and metadata storage requirements included archiving of sensor data streams and of metadata needed to attribute and to report the measurements on both the individual and group levels.

The final system design as implemented is illustrated in Figure 4. It indicates that the POEM system software consists primarily of the edge-node software and the backend server software.

### 2.2.1 Edge Nodes

POEM edge nodes are a user's PCs fitted with ambient sensors and stand-alone plug-level power meters. The original plan was to install power meters on the relevant office power plugs to measure the per-user energy consumption. During system development, a computational method was devised for determining PC energy usage via software based on the power-state occupancy. This eliminated the need for the physical energy meters, and they were installed only for the printers. Details of operation of the energy-management sensor, called SUM vE, and its software implementation are presented in a later section.

The edge-node software, labeled POEM Client in Figure 4, consists of the three functional modules:
- Personal User Interface (UI) and visualization

- Ambient-monitor sensor agent
- Energy sensor and agent, SUM vE

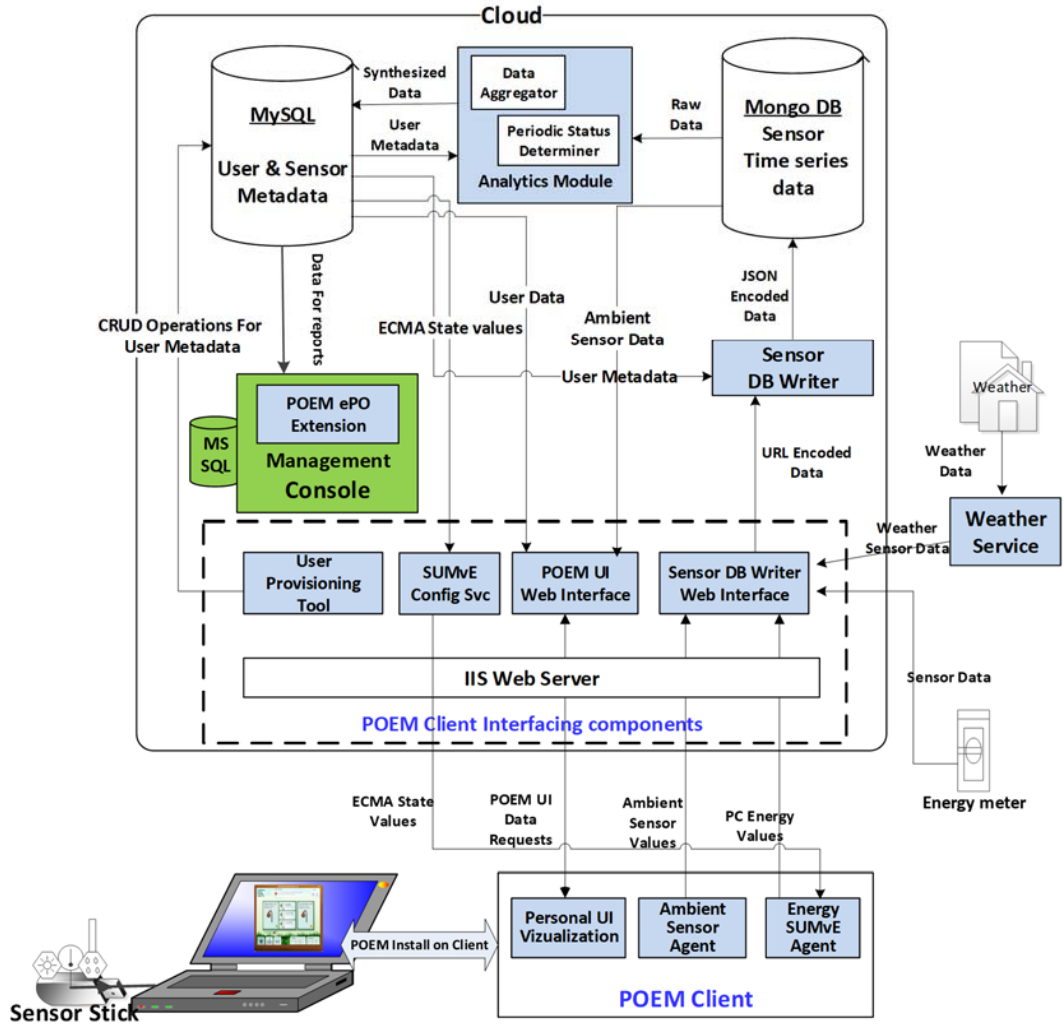

Fig. 4 POEM System Architecture and Implementation

The UI module is in charge of driving the user interface, including the display of relevant data such as depicted in Figures 2 and 3, and supporting the interactions for user inputs. It obtains values to be displayed from the back-end server complex using specified queries.

An ambient-sensor agent communicates with the locally attached USB stick that contains the physical sensors for measuring temperature, humidity and light intensity. The sensor board used for this purpose also includes an embedded microcontroller programmed to implements a subset of USB protocols to communicate with the generic Windows sensor drivers running on the client. The agent builds upon those and implements the interfaces and the logic to read values from the sensors of interest for the subsequent forwarding to the back end.

The energy sensor module contains a service that runs once every second to observe and record the energy state of the machine that it is running on. It uses consecutive and cumulative aggregations of those states to compute the energy usage for the subsequent forwarding to the back end.

### 2.2.2 POEM Back End

The POEM back end is a collection of services that perform various functions, and a few interfacing modules for edge interactions, data ingestion and miscellaneous services.

Some of the major back-end services depicted by the Figure 4, include:
- Sensor database
- User and metadata database
- Analytics module
- Management console

In addition to those, some simpler services provide helper functions. They include a Sensor DB (database) Writer that decouples sensor agents from a specific database implementation, a User Provisioning management tool for adding system users, and a SUM vE configuration service described later.

### 2.2.3 System Flows

A POEM end point becomes operational when the client software is installed, the ambient-sensor stick is inserted in the machine, and the user is provisioned - that is a user ID is assigned and the corresponding static portion of the metadata are entered in databases using the provisioning tool described later. Ambient sensor data are sampled at regular intervals whose length is programmable. They are reported using another programmable interval, in our initial configuration set to 30 s. Energy sensor data are also reported every 30 s. Details of its operation are presented in a separate section on SUM vE.

When the user machine is active, sensor data are posted by the client software in regular intervals using the HTTP post to a dedicated URL that points to an IP and web server application port of the Sensor DB Writer Web Interface in Figure 4. This is an entry point to a web service that acts as the front-end of the sensor database writer helper service. Its function is to write the sensor data in the database and to decouple that operation from the sensor agent. This indirection provides the flexibility to change the database implementation at a later date without impacting sensor agents in the field.

Posted data are humanly readable strings in the rather generic HTML-compliant encoding of name, value pairs in the format that followed the spirit of some related early academic work [13]. A partial example of the POEM reporting serialization format as it appears "on the wire" is illustrated below as the "name":"value" pairs in JSON-style syntax

```
{"id": "userID", "tsutc": "mm-dd-yy-h:mm:ss", "light":50, "tempC:"25, "rh":60, "indicator":"ambsensor"}
```

Names used in the name, value pairs were defined in the POEM vocabulary for consistency in referencing and parsing. The "id" field is used to designate the unique "userID" identifier associated with a PC as generated in the provisioning process. "tsutc" is the time stamp in Coordinated Universal Time (UTC) of the reported sample, and "light", "tempC" and the "rh" are names of the corresponding numeric values of the light, temperature in $^0$C, and relative humidity in %, respectively. Comma is a separator and empty spaces are ignored by the data parser. UTC time was used for consistency across geographies, but the data were converted to local time zones for presentation to users and for analysis.

The example indicates that particular engineering units were implicitly assumed for the ambient sensor reports of the light, temperature and humidity. In retrospect and with the benefit of considerations described in [14], a cleaner design would have been to designate them explicitly via separate metadata fields.

As described, sensor readings are reported as a push operation, as opposed to a pull that would require a specific request from the server complex to each client. No special provisions were made for scalability of data ingestion, since the number of clients in our studies was relatively small – on the order of tens of users – and their aggregate data rates were well within the web server's ability to handle posts. The plan was to support higher aggregate data rates, when needed, by using a load balancer.

In POEM implementation, user ID is mappable and directly correlated with a stream ID which is used as a key for storing and retrieving data streams in the sensor database. Metadata annotations are used to label and to separate the individual data streams emanating from the particular sensors. All other (static) metadata for the user were stored in the relational database in a manner retrievable by the user ID, so they did not need to be reported with individual sensor updates.

In addition to energy and ambient reports, two other sources provided data to the sensor database – power meters and weather service. Commercial power meters had the capability to post data to a programmable IP address in programmable intervals. They were configured to report to a dedicated port in the sensor DB writer web interface module. The program servicing that port parsed the data and forwarded the values of interest, such as the time-stamped current power draw and cumulative energy use, to the sensor database writer.

The weather-service functional module was designed to query a public Internet service, such as the Weather Underground, for the outdoor weather conditions of the specific location where the user building is located. This information was obtained and posted to the database via the sensor DB writer web interface on a regular basis with programmable intervals, typically 15 min. It was retrieved from there for display to the users and was available as an input to BMS if desired.

*2.2.4 Data and Metadata Storage*

Due to the extensive use of metadata and their largely different characteristics and update patterns from the sensor data, a decision was made to split the functionality between two different types of databases. Among other things, this would allow for the separate evolution of data and metadata implementations as understanding of the system improved by experiencing it in operation.

Per-user metadata included designations of user's department, floor and building. Since this portion of user metadata is fairly static and it changes rarely if at all, a design decision was made to store those as tables in a relational database, with a unique user indicator as a key, and other membership fields defining the static schema representing memberships and associations of interest, such as a user's department and floor. In addition, sensor data were annotated with metadata to indicate the type of measurement being reported, such as the temperature, humidity, light and energy. The primary user's office was included as their location in the static metadata since we could not find a suitable indoor location technology that worked in all pilot locations. The alternative of having the users input their location as they moved around with their laptops was deemed to be too imposing in terms of making the POEM application too chatty and potentially annoying.

Metadata storage, due to its mostly associative relations and relatively static nature once defined, was delegated to a traditional relational database. In the implementation described here, MySQL [15] was chosen because it met the requirements and had a license-free open-source version which was preferable for a pilot. Its potential transactional limitations on throughput were not of great concern given the volume and access patterns of metadata (predominantly reads) that were anticipated.

The sensor database implementation requirement was the ability to store streaming data without imposing a strict schema on the data, since we had several different types of sensors – hardware, software, and human input – that we expected to grow by adding new types of sensors to be defined over time. For those reasons and others described in [14], a document NoSQL database was chosen, in this case Mongo DB [16]. The system used unique identifiers that served as keys in both databases so that the related entries could be directly correlated. For example, the relational metadata database used this key to identify the data streams and records in the sensor database related to a specific user.

From a practical point of view, a single database for sensor data and metadata may have been preferable, but we could not find any that would meet our requirements at the time when the project implementation begun.

*2.2.5 Analytics*

The analytics module operated periodically to compute group averages of interest. Every 15 min it computed individual usage trends based on the power consumption in the last cycle compared to the consumption for the corresponding part of the previous day or week average. The corresponding trending indicator is shown as a large colored dot (green) in the lower right corner of the flower images in Figure 2. Green color was used to indicate good trending, red for bad, and orange for marginal. Results were stored in the relational database and made available for display in individual POEM UI screens of the users when requested. Daily averages for all users were computed at midnight and made available to the UI as historical individual and group averages of interest to be displayed as illustrated in Figure 3.

The analytics module was operated in 15 min and in daily intervals to compute aggregate data of interest and to store them in the database for user queries and archival purposes. It retrieved raw data from the time series in the sensor database and computed data for the relevant associations and groups, such as departments, floors and total building data.

Individual user data are furnished to UI portions of individual active users. The POEM UI, when the application is active and displayed on the user's screen, requests data from the POEM UI Web Interface service. It in turn queries the two databases to obtain relevant data and comparisons and furnishes them to the UI. The reverse data flow, from the user to the database, is supported for users to provide feedback on their subjective feeling of comfort.

*2.2.6 Management Console*

The only major remaining system function depicted in Figure 4. is the management-console interface. It was designed to provide the reports of interest to system managers, such as the IT and facilities (building manager). Reports included energy usage per user, department, floor and building. Results of the pervasive ambient sensing were made available to the building manager in order to allow identification of rogue zones that were too hot or cold. Depending on their density, this information may or may not be captured by the BMS sensors, such as thermostats. Moreover, personal ambient sensors provided an independent corroboration of data. Significant discrepancies between the two systems, if encountered, can be used to identify malfunctions in either.

Implementation of the management console and visualization can be a fairly complex task. In order to save the development time and cost, we opted to use the configurable reporting feature of the existing management console used by the IT department for management of security and antivirus functions. It was relatively easy to interface to sensor and metadata database queries that in turn drove the visualization in terms of customary graphs and bar charts. It also eliminated the need for introduction of another dedicated console that system operators usually resent.

*2.2.7 System Energy Sensor, SUM vE*

This section describes some implementation details of a software agent designed to measure the PC energy consumption. It is included as an illustration of sensors that can quantify physical phenomena by inference from observing related indicators that are easier or cheaper to measure. Readers not interested in the details of SUM vE operation may omit this section without the loss of continuity.

The original design intent was to use the plug-load power meters to measure PC energy consumption. In the course of research and gathering of preliminary data, the design team discovered a mechanism that can be used to accurately estimate power consumption by software. The insight was that PCs have a small number of distinct power states in which their energy consumption is fairly constant. The idea came from the ECMA-383 standard [12] that defines a methodology for measuring the energy of personal computing products that is used by the Energy Star program [10]. It defines the following expression for estimating the total annual energy usage of a PC

$$TEC_{estimate} = (8760/1000) * [P_{off} * T_{off} + P_{sleep} * T_{sleep} + P_{idle} * T_{idle} + P_{sidle} * (T_{sidle} + T_{work})]$$

Where TEC is the total energy consumption estimate, $P_i$ is the power consumption in a particular power state and $T_i$ is the fraction of the time that the system typically spends in that state. Power states of interest are "off", "sleep", "idle" when the system is on but the screen is off, "sidle" which stands for the short idle when the system is on and the screen is on, and "work" when the system is in the active mode, i.e. executing a workload, and the screen is on. The multiplier at the beginning comes from the number of hours in a non-leap year, 8760, divided by 1000 to express the energy in kWh.

Power state occupancy is detectable by software through a Windows system call. The energy spent by each power state is measurable or generally available from the manufacturers in their Energy Star specification. The POEM team modified an internally developed system utilization monitor (SUM) that collected PC usage statistics to detect and record power state occupancy. It ran every second and recorded the system state, such as idle or short idle. The energy monitoring agent accumulated the state occupancy over its reporting time and it was then multiplied by the power consumption in each recorded state. That is practically a constant value that can be determined by using the external power meters to measure the power draw of the system in that particular state.

The SUM program tracked the length of occupancy of idle and short idle states. Sleep state duration was derived from the system event log which records the times when going to sleep and waking up or booting. Our own studies indicated that there is virtually no difference in the power drawn between the short idle and the working states. This allowed us to use a single power measure for both, and to avoid the problem of accounting for different workloads. A similar conclusion was reached by the ECMA and the Energy Star who also substituted the working state with short idle for similar reasons.

The SUM vE agent reports data in a similar "name":"value" format as the ambient sensors shown earlier, with user ID for correlation, time stamp, energy data, and a metadata indicator of the sensor type.

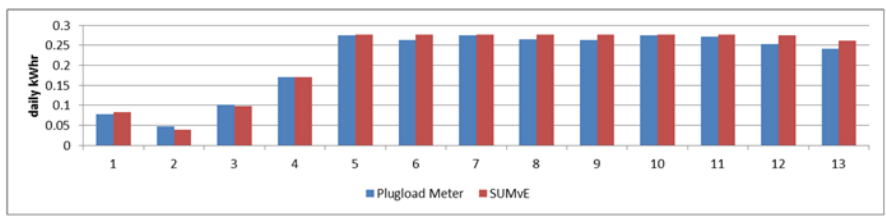

Fig. 5 Energy Monitor Accuracy, SUM vE Compared With Plug-Load Meter

Figure 5 illustrates the results of validation of energy sensor accuracy performed on several machines in a typical office use by comparing the SUM vE computation to the actual consumption measured by an external plug-load meter to which the power cord was connected.

Some additional tweaks involved in completing the SUM vE (version Energy) implementation included adjustments for the external monitors and the laptop battery charging. External monitors are used by all desktops and some laptop users, typically when docked in their office. Monitors are on standby in the long idle and on in the short idle states, so the total systems power consumption may be computed by using their respective on/sleep/off power ratings or measures. For the monitor power tracking, their type designation needed to be included in the user information during system provisioning.

Since the measurement of interest was the PC energy consumed from the building's power distribution network, SUM vE did not report energy consumption for laptops while running on the battery power. However, it did measure the state of battery charge when the system is plugged in, to account for the significantly larger power draw of laptops when charging batteries. This is typically on the order of 2 to 3 times higher than the consumption when plugged in with the battery fully charged.

The use of SUM vE for energy quantification and reporting simplified deployments of the system and saved the added cost of acquiring power meters.

## 2.3 POEM System Deployment and Provisioning

A number of versions of the system were constructed and used for experimentation, development and testing. The final version depicted in Figure 4 was deployed in the two pilot studies with actual office users whose results are described in the subsequent section.

Deployments consisted of the edge nodes represented by the user's office PCs fitted with the POEM client and equipped with ambient sensors connected via a USB port. The server part was designed to run either on premises or in the cloud. Our initial implementation was on the servers dedicated for development and testing purposes. We used virtualized environments on local servers to facilitate porting to a commercial cloud that was completed later. Physical servers were initially located in the research laboratory and were later moved to a professional hosting facility for the benefits of professional IT management and Internet addressability that was necessary for implementing pilots on external corporate networks. A cloud version was later installed and operated using a commercial cloud provider.

Edge and end-user system provisioning process consisted of installing POEM hardware and clients on the participating machines and entering user information in the metadata database. We developed a user provision tool depicted in Figure 4 to simplify the process. Per-user data included user's system ID, office location, department, floor and building. Additional information included power-state draws for idle and short idle for the user's particular machine type. This was used to compute the energy consumption by the SUM vE energy sensor, as indicated by the query line labeled ECMA State Values in Fig 4.

# 3 PILOT DEPLOYMENTS AND USER STUDY

The POEM system was piloted with the actual office users in two separate locations for approximately three months of continuous use. This was followed by a user study and extensive participant interviews performed by the human-factors experts to assess the user experience and feedback on the select system features. The overall goal was to explore the ways of engaging users to actively participate and manage their behaviors in ways that might capture the projected 12% energy savings in operation of energy-efficient buildings with a view towards net-positive energy buildings.

More tangible specific objectives of the pilot deployments were to:
- Close the loop between building occupants and management systems
- Explore methods that might increase user engagement in building operation and energy usage through control of ambient conditions
- Explore the value of pervasive ambient condition sensing and interactive user feedback on personal comfort to building managers and management systems
- Quantify energy usage of IT equipment in office buildings for design and operation of net-positive energy buildings

## 3.1 Pilot Settings and Data Collection

The pilots were conducted for approximately three months with 27 users (regular office workers) in France during the spring, and 46 in Japan during the heating season. Data were collected for the individuals, and aggregates and averages were compiled for the departments and floors. Personal data were provided to individual users via the POEM UI and to building and facilities managers via the management console.

As an illustration of some of the collected data, individual daily readings of temperature and humidity are provided in Figure 6 and Figure 7, respectively. Different days are represented by different colors. Mean laptop energy consumption by the department for one pilot is shown in Figure 8.

Humidity is important as it, coupled with the temperature, impacts the perception of user comfort. Its indoor extremes can create a feeling of extremely damp or dry environment. Most of the readings in the two pilots were within the recommended range of 30% - 60%. This provided a quantified confirmation that BMS systems were operating in the desired range.

Light measurements were collected for reference and potential future control uses.

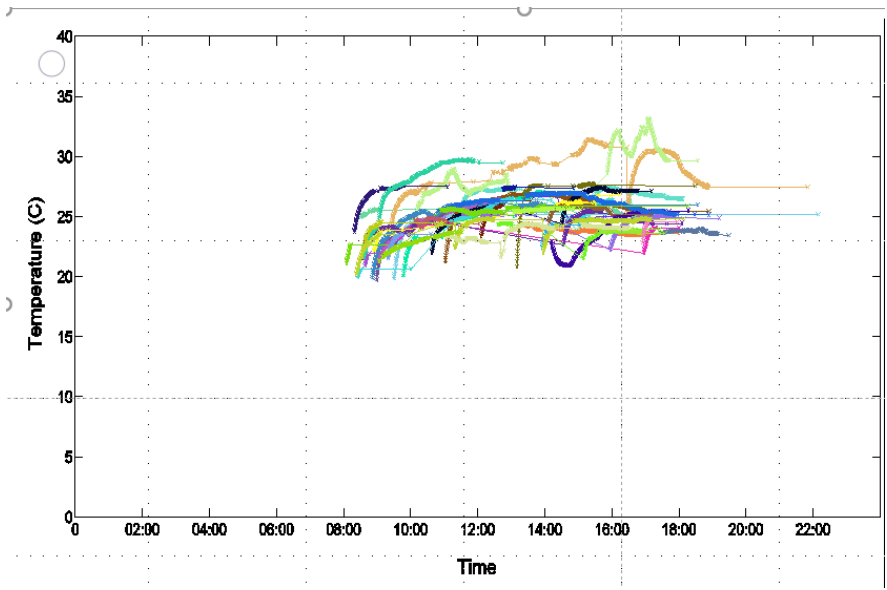

Fig. 6 Personal Ambient Temperature, Daily Readings for an Individual User

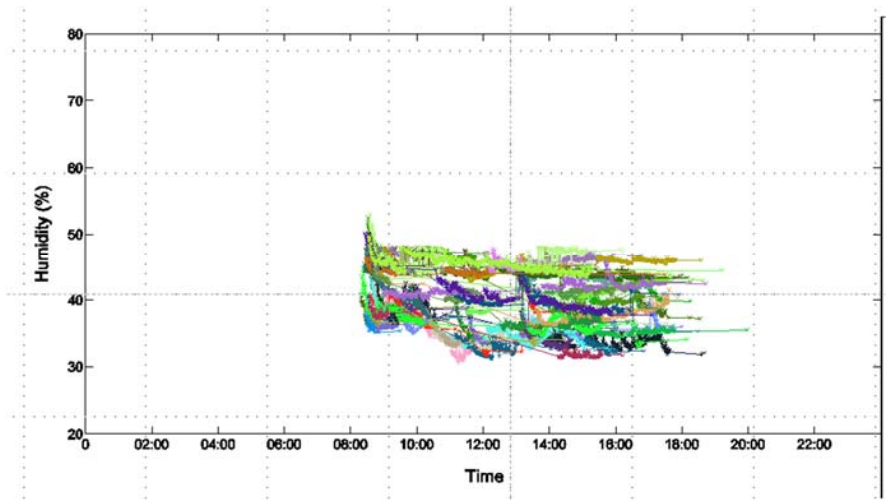

Fig. 7 Personal Ambient Humidity, Daily Readings for an Individual User

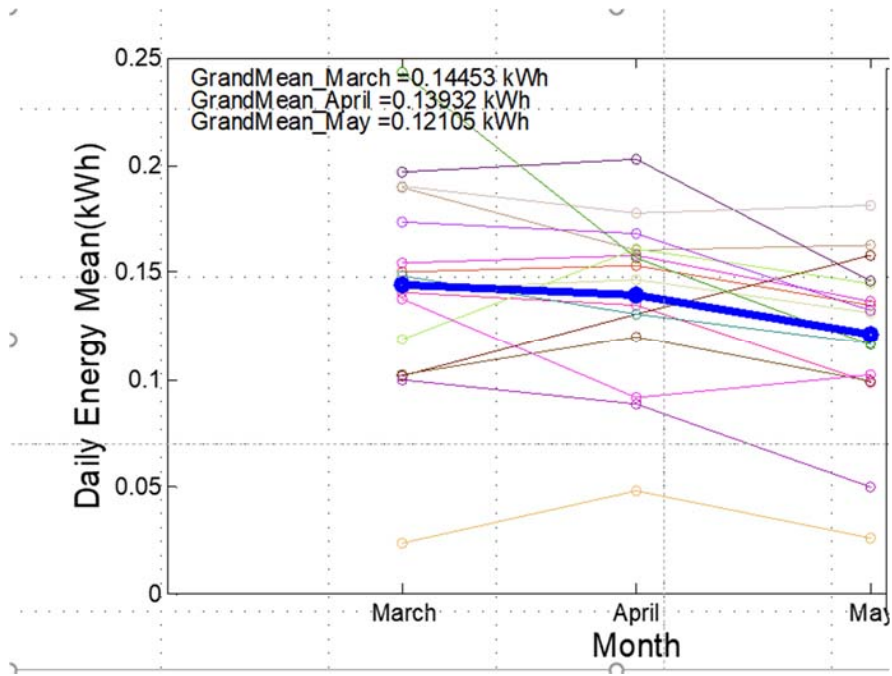

Fig. 8 Mean PC Energy Consumption, by Department

## 3.2 Pilot User Study

Following completion of the pilots, extensive exit interviews were conducted to understand user experiences and their perceived value of system features. Some of the major findings were that users were very interested in having the personal ambient data and energy usage data, especially when coupled with guidance on how to manage them. Ambient data was regarded by users as very useful for organizational culture and they felt that they contributed to creating better working conditions and might improve productivity and effectiveness. Many users expressed the desire to contribute to helping the environment by managing their use of energy in the workplace and were frustrated by the prior lack of metrics and tools to do something about it.

Table 1 shows ranking by users in France of key POEM features in terms of perceived value. Personalized ambient sensing was perceived as the most valuable. Energy consumption came in second and personalized comfort feedback the third. It was later discovered that this was probably due to the inadequate training resulting in users not knowing how to use this pareticular feature.

Comparable responses from users in Japan are presented Table 2. They ranked personal ambient sensing as the most valuable, followed by the comfort feedback and energy consumption. It is unreasonable to draw major conclusions based on the relatively small sample sizes, but the intuitive hypothesis was to attribute these to cultural differences. Answers to some additional questions indicated that users in Japan valued the energy data as well but were frustrated by not knowing what to do about it and how to change them.

Pilot participants felt strongly that the ambient information may contribute to their health and well-being in a work environment. The ability to change their environment was also highly valued. From the point of view of the consortium that initiated the study, these are encouraging results. For the construction companies and office-space lenders, this translates to the increased value and competitiveness of user-friendly spaces. It also allows companies to demonstrate by actions that they care for the comfort and well-being of their employees in the workspace – always a good thing to do and a potential competitive advantage.

Table 1 User Rankings of POEM Features, France

**15. Please rank the following POEM features in order of priority: (1 = most valuable/used; 3= least valuable/used)**

|  | 1 | 2 | 3 | Rating Average | Response Count |
|---|---|---|---|---|---|
| Energy Consumption | 31.6% (6) | 36.8% (7) | 31.6% (6) | 2.00 | 19 |
| Humidity, Temperature, Light Data | 47.4% (9) | 26.3% (5) | 26.3% (5) | 1.79 | 19 |
| Comfort Feedback Tool | 21.1% (4) | 36.8% (7) | 42.1% (8) | 2.21 | 19 |
|  |  |  | answered question |  | 19 |
|  |  |  | skipped question |  | 0 |

Table 2 User Ranking of POEM Features, Japan

**15. Please rank the following POEM features in order of priority: (1 = most valuable/used; 3= least valuable/used)**

|  | 1 | 2 | 3 | Rating Average | Response Count |
|---|---|---|---|---|---|
| Energy Consumption | 16.1% (5) | 19.4% (6) | 64.5% (20) | 2.48 | 31 |
| Humidity, Temperature, Light Data | 48.4% (15) | 38.7% (12) | 12.9% (4) | 1.65 | 31 |
| Comfort Feedback Tool | 35.5% (11) | 41.9% (13) | 22.6% (7) | 1.87 | 31 |
|  |  |  | answered question |  | 31 |
|  |  |  | skipped question |  | 1 |

Aggregate data on energy usage provided a quantification input as a reference to building designers in the consortium. Data reports on the ambient conditions were highly valued by building managers. They provided an independent confirmation of the BMS-reported values and provided much more pervasive localized information that uncovered some problematic zones in buildings that tended to get too hot or too cold at times, due to the window orientation and sun exposure. After the initial hesitance to make the building data available to users, building managers found it extremely valuable to have the personal comfort feedback information and "being able to manage conditions at the level of a human being, as opposed to the impersonal facility management".

On the compliance side, the system helped with compiling reports on employee health conditions that are required in France and energy use and annual reporting on energy uses and savings that are required in Japan.

The final encouragement came at the end of the pilot studies, when many users were lobbying to have the system remain operational and did not want to remove it from their machines.

## 4 IMPLEMENTATION NOTES

The previous sections described the POEM project as a complete example of a sizable IoT installation designed to collect targeted data, carry out planned experiments, validate some hypotheses, and provide quantified insights.

In the creation of POEM, the conceptual design and implementation phases were clearly separated, admittedly perhaps less attributable to virtue than to serendipity. Namely, the team first designed the concept and created an interactive mockup of the UI using a software tool to sell the idea and the concept for funding. Following the approval, it took a team of researchers and developers a better part of a year to actually reduce the concept to practice and to implement the system [17].

Some of the features that were initially anticipated turned out not to be workable or worthwhile and were excluded from the implementation. They included indoor location, plug-load disambiguation, electric vehicle charging and print tracking per user as additional components of the more comprehensive quantification of a user's impact on the environment.

Indoor location was intended to localize tracking of the ambient conditions as users are moving around the office space. We found no reliable commercial or stable research solution that worked in all pilot settings. The fallback position was to install the ambient sensors near the user's desks and in the docking stations and to not have them report the data when their host machine moved.

Plug-load disambiguation works by analyzing the power signatures of measurements produced by the power meters to determine which particular devices on their network are turning on or off. This was meant to indicate when personal office lamps, printers or heaters were on or off. The research available at the time did not provide a reliable way to accomplish this, so the decision was made to omit the feature.

Tracking of the electric vehicle charging was dropped since the buildings in which the pilots were conducted had either very few charging stations or no parking lots. Consequently, the effort required to provide this feature was deemed to be not worthwhile in the view of the miniscule potential benefit to the project.

Only one of the office environments in which the pilots were located had a printing solution that allowed for accounting of individually attributable printer usage. Our initial studies indicated that users were interested in this data, especially with the POEM feature of providing comparisons with the user-relevant averages as was done in the case of energy usage. However, interfacing to this capability required access to the user directory and authentication mechanism, and we did not obtain the permission from the management and the IT department to do so.

Collection of the light data was inspired by a study that indicated that user-directed lighting based on their needs and availability of natural light can improve user satisfaction and save energy [18]. However, we did not have a convenient setting to replicate this capability in the facilities where the pilots were conducted, so the light control was deferred to future versions.

These changes were project specific and are not necessarily applicable to other IoT systems. Their descriptions were included to illustrate that changes based on additional insights can and often do occur between the project conceptualization, implementation, and ultimately operational phases. IoT systems designers and implementation teams should plan for and adapt to these accordingly.

## Acknowledgments

Major contributors to the POEM project include (from Intel) Yves Aillerie, Ulf Hanebutte, Catherine Huang, Sailaja Parthasarathy, Han Pham, Sylvain Sauty, Scott Shull, and Jun Takei. Executive support from Lorie Wigle (Intel), Marie Annick Le Bars (Bouygues Immobilier), and Renaud Deschamps (Lexmark) provided encouragement and funding.